\documentclass[aps,prb,twocolumn,groupedaddress, longbibliography, floatfix]{revtex4-2}
\usepackage[]{graphicx}
\usepackage{braket}
\usepackage{amsmath}%

\begin{document}
\renewcommand{\vec}[1]{\mathbf{#1}}


\title{Exact factorization of the many-body Green's function theory of electrons and nuclei}
\author{Ville J. H\"{a}rk\"{o}nen}
\email[]{ville.j.harkonen@gmail.com}
\affiliation{Computational Physics Laboratory, Tampere University, P.O. Box 692, FI-33014 Tampere, Finland}
\date{\today}

\begin{abstract}
We combine the recently developed many-body Green's function theory for electrons and nuclei with the exact factorization of the wave function. The existing Born-Oppenheimer Green's functions are shown to be special cases of our exact approach. We consider the limitations of the laboratory frame formulation of the Green's function theory and discuss why the body-fixed frame formulation is needed in order to go beyond the Born-Oppenheimer theory. We give exact forms of the electronic and nuclear Green's functions written in terms of the exact factorized states, providing a systematic approach beyond the Born-Oppenheimer approximation. The lowest order approximation to the exact electronic Green's function is found to be an expected value of the Born-Oppenheimer electronic Green's function with respect to the nuclear density.
\end{abstract}

\pacs{63.20.kg}
\keywords{Born-Oppenheimer approximation, Many-body Green's functions, Lattice dynamics, Phonon}

\maketitle


\section{Introduction} 
\label{cha:Introduction}

The Born-Oppenheimer (BO) approximation \cite{Born-Oppenheimer-Adiabatic-Approx.1927,born-huang-dynamical-1954} is the corner stone of our current understanding about the non-relativistic many-body quantum mechanical systems composed of electrons and nuclei. Its validity originates from the mass difference between the electrons and nuclei. There are, however, systems where the validity of the BO approximation is compromised. Examples of such systems are various molecules with conical intersections \cite{Yarkony-DiabolicalConicalIntersections-RevModPhys.68.985-1996,Agostini-WhenTheExactFactorizationMeetsConicalIntersections-2018,Fabri-BornOppenheimerApproximationInOpticalCavitiesFromSuccessToBreakdown-2021}, graphene \cite{Pisana-BreakdownOfTheAdiabaticBornOppenhApproximationInGraphene-2007} and possibly the recently discovered superconductive hydrides \cite{Somayazulu-EvidenceForSuperconductivityAbove260KInLanthanumSuperhydrideAtMegabarPressures-PhysRevLett.122.027001-2019,Drozdov-SuperconductivityAt250KInLanthanumHydrideUnderHighPressures-2019,Flores-APredictionForHotSuperconductivity-2019,Sun-RouteToaSuperconductingPhaseAboveRoomTemperatureInElectronDopedHydrideCompoundsUnderHighPressure-PhysRevLett.123.097001-2019,Snider-RoomTemperatureSuperconductivityInACarbonaceousSulfurHydride-2020,Errea-QuantumCrystalStructureInThe250kelvinSuperconductingLanthanumHydride-2020,Gidopoulos-EnhancedElectronPhononCouplingNearAnElectronicQuantumPhaseTransition-2022}. Therefore, the beyond-BO theoretical and computational methods are expected to play an important role in understanding the properties of these systems. To describe these kinds of systems, several fully quantum mechanical beyond BO approaches have been developed such as the wave function approach \cite{born-huang-dynamical-1954}, the exact factorization \cite{Gidopoulos-ElectronicNonAdiabaticStates-2005,Gidopoulos-Gross-ElectronicNonAdiabaticStates-2014,Abedi-ExactFactorization-PhysRevLett.105.123002-2010,Requist-ExactFactorizationBasedDFTofElectronPhononSystems-PhysRevB.99.165136-2019,Li-EnergyMomentumAndAngularMomentumTransferBetweenElectronsAndNuclei-PhysRevLett.128.113001-2022,Villaseco-ExactFactorizationAdventuresAPromisingApproachForNonBoundStates-2022}, the density matrix renormalization group approach \cite{Muolo-NuclearElectronicAllParticleDensityMatrixRenormalizationGroup-2020,Feldmann-QuantumProtonEffectsFromDensityMatrixRenormalizationGroupCalculations-2022}, the multicomponent density functional theory \cite{Kreibich-MulticompDFTForElectronsAndNuclei-PhysRevLett.86.2984-2001} and the many-body Green's function theory \cite{Harkonen-ManybodyGreensFunctionTheoryOfElectronsAndNucleiBeyondTheBornOppenheimerApproximation-PhysRevB.101.235153-2020}.

The many-body Green's function theory of electrons and nuclei was introduced already in the 1960s by Baym \cite{Baym-field-1961} and has become a useful theory that has already been used in the actual computations, especially in the electron-phonon related studies \cite{Ponce-EPW-2016,Giustino-ElectronPhononInteractFromFirstPrinc-RevModPhys.89.015003-2017,Ponce-FirstPrinciplesPredictionsOfHallAndDriftMobilitiesInSemiconductors-PhysRevResearch.3.043022-2021}. However, it is expected \cite{Harkonen-ManybodyGreensFunctionTheoryOfElectronsAndNucleiBeyondTheBornOppenheimerApproximation-PhysRevB.101.235153-2020} that the theory of Baym is not useful in its general form, that is, when we truly go beyond the BO approximation. In this work, we combine the exact factorization of the wave function \cite{Gidopoulos-ElectronicNonAdiabaticStates-2005,Gidopoulos-Gross-ElectronicNonAdiabaticStates-2014} with the many-body Green's function theories \cite{Baym-field-1961,Harkonen-ManybodyGreensFunctionTheoryOfElectronsAndNucleiBeyondTheBornOppenheimerApproximation-PhysRevB.101.235153-2020}. This is beneficial for several reasons. Firstly, we can develop a perturbation theory in the nuclear kinetic energy allowing systematic beyond BO approximations. The Green's functions are essentially written with respect to the exact factorized states which can in some cases simplify the actual computations. The preceding simplification is obtained since after the exact factorization, the expected values are taken with respect to states in the nuclear space or in the conditional electronic space only, not with respect to the states in the full electron-nuclear space. This is beneficial as we have gained over the years quite extensive experience how to solve the equations in the nuclear space or conditional electronic space. Techniques to solve the aforementioned equations are implemented, in the special case of the BO approximation, in several existing computational packages. Moreover, the approach taken here provides a way to see more specifically what approximations have to be imposed in order for the laboratory frame formulation of the Green's function theory of Baym to give useful results. We will find out, on the other hand, when the body-fixed approaches \cite{Harkonen-ManybodyGreensFunctionTheoryOfElectronsAndNucleiBeyondTheBornOppenheimerApproximation-PhysRevB.101.235153-2020} are needed. To summarize, our aim here is to establish an exact factorization of the states with respect to which the exact electron and nuclear Green's functions are defined. As a result, we will obtain a systematic and still exact Green's function approach beyond the BO approximation, which will be computationally more accessible in comparison with the non-factorized Green's functions we start with. In the process, we will find out the limitations of the laboratory frame formulation of the Green's function theory.

This paper is organized as follows. We start by recapping the central equations in the laboratory frame exact factorization of the wave function in Sec. \ref{cha:HamiltonianAndBornOppenheimerApproximation} and deduce the BO approximation as a special case. We discuss the symmetry related topics in Sec. \ref{cha:Symmetry} and show that the laboratory frame formulation does not give a reasonable nuclear density, despite being useful in the computation of phonon spectra and related properties. The laboratory and body-fixed frame theories of the electronic and nuclear many-body Green's functions within and beyond the BO approximation are considered in Sec. \ref{cha:GreensFunctionTheories}. We start with the BO Green's function theory in Sec. \ref{cha:WithinBornOppenheimer} and in Sec. \ref{cha:BeyondBornOppenheimer}, introduce the laboratory and body-fixed frame Green's functions. In Sec. \ref{cha:ManyBodyPerturbationTheoryInExactFactorization} (laboratory frame formulation) and in Appendix \ref{PerturbationExpansionInTheBodyFixedFrame} (body-fixed frame formulation), we derive the many-body perturbation theory in the nuclear kinetic energy with respect to exact factorized states. Here we also show how to obtain the BO Green's function theory as a special case of our exact approach.

\section{Preliminaries}
\label{cha:Preliminaries}

\subsection{Hamiltonian and Born-Oppenheimer approximation}
\label{cha:HamiltonianAndBornOppenheimerApproximation}

The object of our study is the system of $N_{e}$ electrons and $N_{n}$ nuclei described by the Hamiltonian of the form
\begin{equation} 
H = T_{e} + T_{n} + V_{ee} + V_{en} + V_{nn},
\label{eq:PreliminariesEq_1}
\end{equation}
where $T_{e}$ is the kinetic energy of electrons, $T_{n}$ is the nuclear kinetic energy and $V_{ee}$, $V_{en}$ and $V_{nn}$ are the electron-electron, electron-nuclei and nuclei-nuclei potential energies of the Coulomb form, respectively. The time-independent Schr\"{o}dinger equation for this Hamiltonian can be written as
\begin{equation} 
H \Psi\left(\vec{r}, \vec{R}\right) = E \Psi\left(\vec{r}, \vec{R}\right),
\label{eq:PreliminariesEq_2}
\end{equation}
where $\vec{r}$ and $\vec{R}$ denote all the electronic and nuclear coordinates, respectively (see Appendix \ref{Notation} for the notation). The wave function is normalized as
\begin{equation} 
\int d\vec{r} \int d\vec{R} \left|\Psi\left(\vec{r}, \vec{R}\right)\right|^{2} = 1.
\label{eq:PreliminariesEq_2_2}
\end{equation}
The exact wave function $\Psi\left(\vec{r}, \vec{R}\right)$ can be written, after the exact factorization \cite{Hunter-ConditionalProbInWaveMech-1975,Gidopoulos-ElectronicNonAdiabaticStates-2005,Gidopoulos-Gross-ElectronicNonAdiabaticStates-2014}, as
\begin{equation}
\Psi\left(\vec{r},\vec{R}\right) = \Phi_{\vec{R}}\left(\vec{r}\right) \chi\left(\vec{R}\right).
\label{eq:PreliminariesEq_3}
\end{equation}
It has been shown that the electronic wave function $\Phi_{\vec{R}}\left(\vec{r}\right)$ and the nuclear wave function $\chi\left(\vec{R}\right)$ in exact factorization satisfy \cite{Gidopoulos-Gross-ElectronicNonAdiabaticStates-2014}
\begin{eqnarray} 
H_{n} \chi\left(\vec{R}\right) &=& E \chi\left(\vec{R}\right), \nonumber \\
H_{e} \Phi_{\vec{R}}\left(\vec{r}\right) &=& \epsilon\left(\vec{R}\right) \Phi_{\vec{R}}\left(\vec{r}\right),
\label{eq:PreliminariesEq_4}
\end{eqnarray}
where the Hamiltonians are
\begin{eqnarray} 
H_{n} &=& \sum_{k} \frac{ 1 }{2M_{k}} \left[-i \nabla_{\vec{R}_{k}} + \vec{A}_{k}\left(\vec{R}\right) \right]^{2} + \epsilon\left(\vec{R}\right), \nonumber \\
H_{e} &=& H_{BO}\left(\vec{r},\vec{R}\right) +  U_{en}\left(\vec{R}\right),
\label{eq:PreliminariesEq_5}
\end{eqnarray}
and
\begin{eqnarray} 
\epsilon\left(\vec{R}\right) &=& \int d\vec{r} \Phi^{\ast}_{\vec{R}}\left(\vec{r}\right) H_{e} \Phi_{\vec{R}}\left(\vec{r}\right), \nonumber \\
\vec{A}_{k}\left(\vec{R}\right) &=& -i \int d\vec{r} \Phi^{\ast}_{\vec{R}}\left(\vec{r}\right)  \nabla_{\vec{R}_{k}} \Phi_{\vec{R}}\left(\vec{r}\right).
\label{eq:PreliminariesEq_6}
\end{eqnarray}
In Eq. \ref{eq:PreliminariesEq_5}, the Born-Oppenheimer (BO) Hamiltonian is defined as $H_{BO} \equiv H - T_{n}$ and the operator $U_{en}$ is acting on the nuclear variables only and is of the form
\begin{eqnarray} 
U_{en}\left(\vec{R}\right) &=& \sum_{k} \frac{ 1 }{2 M_{k} } \left[ \left(-i \nabla_{\vec{R}_{k}} - \vec{A}_{k}\right)^{2} \right. \nonumber \\
&&+ \left. 2 \left( \vec{D}_{k} + \vec{A}_{k}\right) \cdot \left(-i \nabla_{\vec{R}_{k}} - \vec{A}_{k}\right) \right],
\label{eq:PreliminariesEq_6_1}
\end{eqnarray}
where $\vec{D}_{k}\left(\vec{R}\right) = -i \chi^{-1}\left(\vec{R}\right) \nabla_{\vec{R}_{k}} \chi\left(\vec{R}\right)$. The wave functions in exact factorization are normalized as
\begin{equation} 
\int d\vec{R} \left| \chi\left(\vec{R}\right) \right|^{2} = \int d\vec{r} \left|\Phi_{\vec{R}}\left(\vec{r}\right)\right|^{2} = 1.
\label{eq:PreliminariesEq_6_2}
\end{equation}
The solution of the exact factorized equations in Eq. \ref{eq:PreliminariesEq_4} provide an exact and alternative way to solve the original Schr\"{o}dinger equation given by Eq. \ref{eq:PreliminariesEq_2}.

If we approximate $U_{en} \approx 0$ and $\vec{A}_{k} \approx \vec{0}$, Eq. \ref{eq:PreliminariesEq_4} becomes
\begin{eqnarray} 
H_{ph} \chi\left(\vec{R}\right) &=& E \chi\left(\vec{R}\right), \nonumber \\
H_{BO} \Phi_{\vec{R}}\left(\vec{r}\right) &=& \epsilon_{BO}\left(\vec{R}\right) \Phi_{\vec{R}}\left(\vec{r}\right),
\label{eq:PreliminariesEq_7}
\end{eqnarray}
where the nuclear Hamiltonian is of the form
\begin{equation} 
H_{ph} = T_{n} + \epsilon_{BO},
\label{eq:PreliminariesEq_8}
\end{equation}
and is often called the phonon Hamiltonian. The relations in Eq. \ref{eq:PreliminariesEq_7} are exactly the equations for electrons and nuclei in the BO approximation \cite{born-huang-dynamical-1954}. From the electronic BO equation given by Eq. \ref{eq:PreliminariesEq_7} we see that the BO energy has a parametric dependence on the nuclear variables. This holds since all the nuclear variables $\vec{R}_{k}$ commute with every quantity that appears in the Hamiltonian $H_{BO}$ and thus any function is an eigenfunction of $\vec{R}$. Therefore, $\vec{R}$ in the electronic BO equation can be treated as constants \cite{Dirac-PrinciplesOfQM-1958}, or in other words, as parameters. In the nuclear equations of Eqs. \ref{eq:PreliminariesEq_4} and Eq. \ref{eq:PreliminariesEq_7}, on the other hand, the nuclear variables $\vec{R}$ appear as operators. We note that $\Phi_{\vec{R}}\left(\vec{r}\right)$, satisfying Eq. \ref{eq:PreliminariesEq_4} or Eq. \ref{eq:PreliminariesEq_7}, belongs to a different space of functions than $\Psi\left(\vec{r}, \vec{R}\right)$, even though both seem to be eigenfunctions of the Hamiltonians which are functions of the nuclear variables $\vec{R}$.

Much of our understanding about the electronic structure of crystals is based on the electronic BO equation (the second relation of Eq. \ref{eq:PreliminariesEq_7}) while the theory of lattice dynamics \cite{born-huang-dynamical-1954,maradudin-dyn-prop-solids-1974} is based on the first relation of Eq. \ref{eq:PreliminariesEq_7}. The nuclear problem has an exact solution when the BO energy $\epsilon_{BO}\left(\vec{R}\right) = \epsilon_{BO}\left(\vec{x} + \vec{u}\right)$ is expanded to a Taylor series up to second order in the displacements $\vec{u}$ about the reference positions $\vec{x}$ which are treated as parameters. This is called the harmonic approximation and the diagonalization of $H_{ph}$ can be established in terms of normal or phonon coordinates \cite{born-huang-dynamical-1954}, or by using unitary transformations \cite{Harkonen-OnTheDiagonalizationOfQuadraticHamiltonians-2021}.

\subsection{Symmetry}
\label{cha:Symmetry}

The Hamiltonian $H$ given by Eq. \ref{eq:PreliminariesEq_1} is invariant under the translations and rotations of all particle coordinates. It is known that these symmetries render Eq. \ref{eq:PreliminariesEq_2} useless as such \cite{Sutcliffe-TheDecouplingOfElectronicAndNuclearMotions-2000,Harkonen-ManybodyGreensFunctionTheoryOfElectronsAndNucleiBeyondTheBornOppenheimerApproximation-PhysRevB.101.235153-2020}, when we want to describe molecules or crystals. More specifically, it can be shown that $H$ has purely continuous spectrum ($E$ in Eq. \ref{eq:PreliminariesEq_2} continuous) and does not describe bound states, like molecules or solids. For instance, from the translational invariance it follows that the eigenbasis of the Hamiltonian can be chosen, without a loss of generality, to be the plane wave eigenbasis of the total momentum. This in turn, leads to a constant electron density \cite{Harkonen-ManybodyGreensFunctionTheoryOfElectronsAndNucleiBeyondTheBornOppenheimerApproximation-PhysRevB.101.235153-2020}. By looking Eq. \ref{eq:PreliminariesEq_4} or \ref{eq:PreliminariesEq_7} we see that the exact nuclear equation has the same eigenvalue as the original equation in Eq. \ref{eq:PreliminariesEq_2}, implying that neither of these equations will have a discrete spectrum, if not any further assumptions are made. There are at least two possibilities to deal with this inconsistency.

When the BO approximation is applied to crystals, the Hamiltonian $H$ of Eq. \ref{eq:PreliminariesEq_1} is used as such and Eq. \ref{eq:PreliminariesEq_7} follows as an approximation. In the electronic equation, it is noted that the nuclear variables are parameters and can be chosen freely which locates the system in space. Even though the BO Hamiltonian $H_{BO}$ has the same translational and rotational symmetry as the full Hamiltonian $H$, only the symmetry with respect to the electronic variables for fixed nuclear variables is considered. For instance, in deriving Bloch's theorem \cite{Kittel-QuantumTheoryOfSolids-1987}, only the electronic variables are displaced while the nuclear variables remain fixed. The continuous translational symmetry is broken and the system has only discrete translational symmetry having the lattice periodicity determined by the fixed $\vec{R}$. In the theory of lattice dynamics, described by the nuclear equation (Eq. \ref{eq:PreliminariesEq_7}), only the symmetry with respect to $\vec{R}$ is taken into account \cite{maradudin1971-harm-appr}. Namely, the symmetry considerations in these cases are essentially based on the invariance of the BO energy $\epsilon_{BO}\left(\vec{R}\right)$ under the symmetry operations in the crystallographic space groups. This treatment is necessarily approximate since Eq. \ref{eq:PreliminariesEq_2} as such leads to continuous eigenvalues $E$ which also appear in Eq. \ref{eq:PreliminariesEq_7}, but in the theory of lattice dynamics we get a discrete phonon spectra. The spectra are discrete since the Born-von K\'{a}rm\'{a}n boundary conditions \cite{born-huang-dynamical-1954} are imposed and it has been shown that this is a well-justified approximation, provided the generating volume is taken to be sufficiently large. In this approach, the nuclear density is not a useful quantity as justified in \cite{Harkonen-ManybodyGreensFunctionTheoryOfElectronsAndNucleiBeyondTheBornOppenheimerApproximation-PhysRevB.101.235153-2020}. This can be seen by assuming the harmonic approximation in Eq. \ref{eq:PreliminariesEq_7}. The resulting wave function is of the product form where each of the terms is of the simple harmonic oscillator form \cite{Harkonen-OnTheDiagonalizationOfQuadraticHamiltonians-2021}. The lowest energy vibrational modes have zero frequency guaranteed by the so-called acoustic sum rule originating from the translational symmetry \cite{born-huang-dynamical-1954}. This renders $\chi$ and thus $\Psi \approx \chi \Phi_{\vec{R}}$ useless since the nuclear wave function $\chi$ is equal to zero. The method nevertheless gives phonon spectra rather closely resembling those obtained experimentally \cite{Harkonen-NTE-2014} and we can compute various nuclei related physical quantities without using the nuclear wave function in position representation \cite{Harkonen-ElasticConstantsAndThermodynamicalQuantitiesForCrystalLatticesFromManyBodyPerturbationTheory-2016,Harkonen-Tcond-II-VIII-PhysRevB.93.024307-2016,Harkonen-AbInitioComputStudyOnTheLattThermalCondOfZintlClathrates-PhysRevB.94.054310-2016,Beekman-ThermalAndMechanicalPropertiesOfTheClathrateII-PhysRevB.105.214114-2022}. Problems will appear in this approach if we need for instance the nuclear density, as in the coupled set of equations of motion for the electronic and nuclear Green's functions beyond the BO approximation \cite{Giustino-ElectronPhononInteractFromFirstPrinc-RevModPhys.89.015003-2017,Harkonen-ManybodyGreensFunctionTheoryOfElectronsAndNucleiBeyondTheBornOppenheimerApproximation-PhysRevB.101.235153-2020}.

In the case of molecules, the mentioned issues are handled by reformulating the problem in a different frame of reference \cite{Born-Oppenheimer-Adiabatic-Approx.1927,Wilson-MolecularVibrationsTheTheoryOfInfraredAndRamanVibrationalSpectra-1955,Littlejohn-GaugeFieldsInTheSeparOfRotatAndIntMotionsIntheNbodyProb-RevModPhys.69.213-1997,Sutcliffe-TheDecouplingOfElectronicAndNuclearMotions-2000,Harkonen-ManybodyGreensFunctionTheoryOfElectronsAndNucleiBeyondTheBornOppenheimerApproximation-PhysRevB.101.235153-2020}. In this approach, we establish a coordinate transformation to a body-fixed frame and all the observables are written in this different frame of reference. In this case, Eq. \ref{eq:PreliminariesEq_4} or \ref{eq:PreliminariesEq_7} are not precisely the same anymore, have additional terms and are written in terms of the variables in the body-fixed frame. The total Hamiltonian can be written as \cite{Harkonen-ManybodyGreensFunctionTheoryOfElectronsAndNucleiBeyondTheBornOppenheimerApproximation-PhysRevB.101.235153-2020} $H = T_{cm} + H_{b}$, where $T_{cm}$ is the center-of-mass kinetic energy and $H_{b}$ the remaining part of the Hamiltonian. The Hamiltonian $H_{b}$ is written in terms of the body-fixed frame variables describing the internal motion of the system, the rotational degrees of freedom and the coupling of the internal motion with the rotational degrees of freedom. It is important to note that the Hamiltonian written as $H = T_{cm} + H_{b}$ is still exact and thus has the original symmetries relative to the original variables $\vec{r}$ and $\vec{R}$. The continuous translational and rotational symmetries are broken in the Hamiltonian $H_{b}$, but it may still have system dependent discrete symmetries discussed above when the Born-Oppenheimer approximation is imposed.

\section{Green's Function Theories}
\label{cha:GreensFunctionTheories}

\subsection{Within Born-Oppenheimer}
\label{cha:WithinBornOppenheimer}

We start with the many-body Green's function theory in the BO approximation based on Eq. \ref{eq:PreliminariesEq_7}. The electronic degrees of freedom are treated in second quantization while the nuclear variables in first quantization. The BO Hamiltonian $\hat{H}_{BO}$ is a sum of one and two-body electronic operators written in terms of field operators and are acting on states of the form $\ket{\Phi_{\vec{R}}}$ belonging to the electronic Hilbert space and have a parametric dependence on $\vec{R}$. We then define the electronic one-body Green's function as
\begin{equation} 
G^{BO}_{\vec{R}}\left(\vec{y}t,\vec{y}'t'\right) \equiv \frac{1}{i} \frac{\text{Tr}\left[e^{-\beta \hat{H}^{M}_{BO}} \mathcal{T}\left\{ \hat{\psi}\left(\vec{y}t\right) \hat{\psi}^{\dagger}\left(\vec{y}'t'\right) \right\} \right]_{\Phi_{\vec{R}}}}{\text{Tr}\left[e^{-\beta \hat{H}^{M}_{BO}}\right]_{\Phi_{\vec{R}}}},
\label{eq:WithinBornOppenheimerEq_1}
\end{equation}
where $\beta = k^{-1}_{B} T^{-1}$ and $\hat{\psi}\left(\vec{y}t\right) \equiv \hat{U}^{\dagger}_{BO}\left(t\right) \hat{\psi}\left(\vec{y}\right) \hat{U}_{BO}\left(t\right)$ is an operator in the Heisenberg picture, the evolution operator being solved from the electronic equation for the BO Hamiltonian $H_{BO}$. Here, $\hat{H}^{M}_{BO} \equiv \hat{H}_{BO} - \mu_{e} \hat{N}_{e}$, where $\mu_{e}$ is the chemical potential of the electrons, $\hat{N}_{e}$ the electron number operator and $\mathcal{T}\left\{\cdots\right\}$ denotes the time-ordering. In Eq. \ref{eq:WithinBornOppenheimerEq_1} the trace is taken with respect to the Born-Oppenheimer states
\begin{equation} 
\text{Tr}\left[\hat{o}\right]_{\Phi_{\vec{R}}} = \sum_{m} \braket{\Phi^{\left(m\right)}_{\vec{R}}|\hat{o}|\Phi^{\left(m\right)}_{\vec{R}}},
\label{eq:WithinBornOppenheimerEq_2}
\end{equation}
where $m$ labels the electronic BO states and $\hat{o}$ is an operator acting in the electronic Hilbert space. We assume that $\hat{o}$ is independent of $\vec{R}$. We note that the trace in Eq. \ref{eq:WithinBornOppenheimerEq_2}, and thus the Green's function in Eq. \ref{eq:WithinBornOppenheimerEq_1}, are dependent on the nuclear variables $\vec{R}$. The theory of Green's function $G^{BO}_{\vec{R}}\left(\vec{y}t,\vec{y}'t'\right)$ is well-known, has been discussed extensively in the literature \cite{Kadanoff-Baym-Quant.Stat.Mech-1962,Gross-ManyParticleTheory-1991,Stefanucci-Leeuwen-many-body-book-2013} and has also become a valuable computational tool in the description of realistic materials \cite{Golze-TheGWcompendiumApracticalGuideToTheoreticalPhotoemissionSpectroscopy-2019}. Two main approaches are used to solve $G^{BO}_{\vec{R}}\left(\vec{y}t,\vec{y}'t'\right)$, namely, the many-body perturbation theory and the equations of motion. Both ways provide exact results for the electronic BO problem, but in practice approximations are needed.

The many-body Green's function theory for the nuclei in the BO approximation \cite{Maradudin-Fein-PhysRev.128.2589-Scat-Neutr.1962} is based on the nuclear equation of Eq. \ref{eq:PreliminariesEq_7} with the Hamiltonian given by Eq. \ref{eq:PreliminariesEq_8}. As in the wave function approach, we write $\vec{R} = \vec{x} + \vec{u}$ and expand the potential $E^{BO}_{m}\left(\vec{x} + \vec{u}\right)$ in $\vec{u}$ about the parameters $\vec{x}$. The nuclear Green's functions are then defined as
\begin{equation} 
D^{BO}_{\alpha_{\bar{n}}}\left(k_{\bar{n}}t_{\bar{n}}\right) \equiv \frac{1}{i^{n-1}} \frac{\text{Tr}\left[e^{-\beta \hat{H}_{ph}} \mathcal{T}\left\{ \hat{u}_{\alpha_{\bar{n}}}\left(k_{\bar{n}}t_{\bar{n}}\right) \right\} \right]_{\chi}}{\text{Tr}\left[e^{-\beta \hat{H}_{ph}}\right]_{\chi}},
\label{eq:WithinBornOppenheimerEq_3}
\end{equation}
where the trace is taken with respect to the states $\ket{\chi}$ in the nuclear space. Moreover, we define the notation used for $\hat{u}_{\alpha_{\bar{n}}}\left(k_{\bar{n}}t_{\bar{n}}\right)$ in Eq. \ref{eq:NotationEq_2} of Appendix \ref{Notation}. Here the operators like $\hat{u}_{\alpha}\left(kt\right) = \hat{U}^{\dagger}_{ph}\left(t\right) \hat{u}_{\alpha}\left(k\right) \hat{U}_{ph}\left(t\right)$ are operators in the Heisenberg picture and the evolution operator is written for the Hamiltonian $\hat{H}_{ph}$. If the Hamiltonian $\hat{H}_{ph}$ is used as such, without making any further transformations, for instance to the phonon coordinates, also the momentum functions of the form given by Eq. \ref{eq:WithinBornOppenheimerEq_3} are needed up to $n = 2$ in order to compute the total energy.

The many-body Green's functions defined by Eq. \ref{eq:WithinBornOppenheimerEq_1} and \ref{eq:WithinBornOppenheimerEq_3} together with the nuclear momentum functions, can be used to compute arbitrary one- or two-body electronic observable and $n$-body nuclear observable, including the total energy of the system. These functions can also be used to determine the electronic structure and the vibrational states of the system. These quantities form a complete, but approximate theory to the general problem corresponding to Eqs. \ref{eq:PreliminariesEq_1} and \ref{eq:PreliminariesEq_2}.

\subsection{Beyond Born-Oppenheimer}
\label{cha:BeyondBornOppenheimer}

The beyond-BO theory of many-body Green's functions assumes the Hamiltonian $H$ of Eq. \ref{eq:PreliminariesEq_1} as a starting point. The electronic operators are written in terms of field operators while the nuclear operators in first quantization. The coordinate transformation $\vec{R} = \vec{x} + \vec{u}$ is usually, but not always \cite{Gillis-SelfConsistentPhononsAndTheCoupledElectronPhononSystem-PhysRevB.1.1872-1970}, established at this point as well as the expansion up to second order in $\vec{u}$ \cite{Baym-field-1961,Giustino-ElectronPhononInteractFromFirstPrinc-RevModPhys.89.015003-2017}. The many-body Green's functions are then defined as
\begin{equation} 
G\left(\vec{y}t,\vec{y}'t'\right) \equiv \frac{1}{i} \frac{\text{Tr}\left[e^{-\beta \hat{H}^{M}} \mathcal{T}\left\{ \hat{\psi}\left(\vec{y}t\right) \hat{\psi}^{\dagger}\left(\vec{y}'t'\right) \right\} \right]_{\Psi}}{\text{Tr}\left[e^{-\beta \hat{H}^{M}}\right]_{\Psi}},
\label{eq:BeyondBornOppenheimerEq_1}
\end{equation}
and
\begin{equation} 
D_{\alpha_{\bar{n}}}\left(k_{\bar{n}}t_{\bar{n}}\right) \equiv \frac{1}{i^{n-1}} \frac{\text{Tr}\left[e^{-\beta \hat{H}^{M}} \mathcal{T}\left\{ \hat{u}_{\alpha_{\bar{n}}}\left(k_{\bar{n}}t_{\bar{n}}\right) \right\} \right]_{\Psi}}{\text{Tr}\left[e^{-\beta \hat{H}^{M}}\right]_{\Psi}}.
\label{eq:BeyondBornOppenheimerEq_2}
\end{equation}
The trace in Eqs. \ref{eq:BeyondBornOppenheimerEq_1} and \ref{eq:BeyondBornOppenheimerEq_2} is of the form
\begin{equation} 
\text{Tr}\left[\hat{o}\right]_{\Psi} = \sum_{m} \braket{\Psi_{m}|\hat{o}|\Psi_{m}},
\label{eq:BeyondBornOppenheimerEq_2_2}
\end{equation}
where $\hat{o}$ is an operator acting in the full electron-nuclear space to which the states $\ket{\Psi_{m}}$ belong. In Eq. \ref{eq:BeyondBornOppenheimerEq_2_2}, $m$ labels the eigenstates of Eq. \ref{eq:PreliminariesEq_2}. Strictly speaking, as discussed in Sec. \ref{cha:Symmetry}, the eigenstates of Eq. \ref{eq:PreliminariesEq_2} cannot be labeled with discrete quantum numbers. Since we are discussing crystals, we have imposed periodic boundary conditions such that the spectrum will be discrete and the trace in Eq. \ref{eq:BeyondBornOppenheimerEq_2_2} can be written as a sum. The definitions of $G\left(\vec{y}t,\vec{y}'t'\right)$ and $D_{\alpha_{\bar{n}}}\left(k_{\bar{n}}t_{\bar{n}}\right)$ above resemble to Eqs. \ref{eq:WithinBornOppenheimerEq_1} and \ref{eq:WithinBornOppenheimerEq_3} except that we have replaced the Hamiltonians $\hat{H}_{BO}$ and $\hat{H}_{ph}$ by the Hamiltonian $\hat{H}$ in the ensemble averages and in the evolution operators. We emphasize that the traces in Eqs. \ref{eq:BeyondBornOppenheimerEq_1} and \ref{eq:BeyondBornOppenheimerEq_2} are taken with respect to states that belong to the full electron-nuclear space, not only to the electron or nuclear space, as in Eqs. \ref{eq:WithinBornOppenheimerEq_1} and \ref{eq:WithinBornOppenheimerEq_3}. The Hamiltonian is still exact, and we have not expanded the Hamiltonian in $\hat{\vec{u}}$ at this point. Thus, in Eqs. \ref{eq:BeyondBornOppenheimerEq_1} and \ref{eq:BeyondBornOppenheimerEq_2} the operators in the Heisenberg picture are defined $\hat{\psi}\left(\vec{y}t\right) \equiv \hat{U}^{\dagger}\left(t\right) \hat{\psi}\left(\vec{y}\right) \hat{U}\left(t\right)$ and $\hat{u}_{\alpha}\left(kt\right) = \hat{U}^{\dagger}\left(t\right) \hat{u}_{\alpha}\left(k\right) \hat{U}\left(t\right)$, where the evolution operator $\hat{U}\left(t\right)$ is written for the Hamiltonian $\hat{H}$. The equations of motion for $D_{\alpha_{\bar{2}}}\left(k_{\bar{2}}t_{\bar{2}}\right)$ was first derived by Baym \cite{Baym-field-1961}. The equations of motion for the electronic and nuclear Green's functions, when the electronic equations are written in the Hedin's equation form \cite{Hedin-NewMethForCalcTheOneParticleGreensFunctWithApplToTheElectronGasProb-PhysRev.139.A796-1965}, are called the Hedin-Baym equations \cite{Giustino-ElectronPhononInteractFromFirstPrinc-RevModPhys.89.015003-2017}. The theory based on Eqs. \ref{eq:BeyondBornOppenheimerEq_1} and \ref{eq:BeyondBornOppenheimerEq_2} is exact, but as our discussion of Sec. \ref{cha:Symmetry} implies, these Green's functions as such do not describe bound states. At the same time, the Hedin-Baym equations are the state-of-the-art in describing electron-phonon interactions and related effects in crystals \cite{Ponce-EPW-2016,Giustino-ElectronPhononInteractFromFirstPrinc-RevModPhys.89.015003-2017,Ponce-FirstPrinciplesPredictionsOfHallAndDriftMobilitiesInSemiconductors-PhysRevResearch.3.043022-2021}. In Sec. \ref{cha:ManyBodyPerturbationTheoryInExactFactorization} we go through what approximations have to be made in Eqs. \ref{eq:BeyondBornOppenheimerEq_1} and \ref{eq:BeyondBornOppenheimerEq_2} in order to obtain physically relevant results from these Green's functions and thus from the Hedin-Baym equations.

Next we make a connection between the Green's functions in the laboratory frame given by Eqs. \ref{eq:BeyondBornOppenheimerEq_1} and \ref{eq:BeyondBornOppenheimerEq_2} and the body-fixed Green's functions introduced in Ref. \cite{Harkonen-ManybodyGreensFunctionTheoryOfElectronsAndNucleiBeyondTheBornOppenheimerApproximation-PhysRevB.101.235153-2020}. As in Sec. \ref{cha:Symmetry}, we write the Hamiltonian as $\hat{H} = \hat{T}_{cm} + \hat{H}_{b}$, see Eqs. 16-18 of Ref. \cite{Harkonen-ManybodyGreensFunctionTheoryOfElectronsAndNucleiBeyondTheBornOppenheimerApproximation-PhysRevB.101.235153-2020}. We note that $\hat{T}_{cm}$ commutes with $\hat{H}_{b}$ and thus with all the operators included, like the electronic field operators $\hat{\phi}\left(\vec{y}t\right), \hat{\phi}^{\dagger}\left(\vec{y}'t'\right)$ or the nuclear variables $\hat{\vec{R}}' = \vec{x}' + \hat{\vec{u}}'$, which here refer to the Hamiltonian in the body-fixed frame. Therefore, the exact states are of the product form $\ket{\Psi} = \ket{\phi_{cm}} \otimes \ket{\Psi_{b}}$ and we can trace out the center-of-mass kinetic energy terms from the Green's functions. We define the Green's functions in the body-fixed frame for the body-fixed variables as we did in Eqs. \ref{eq:BeyondBornOppenheimerEq_1} and \ref{eq:BeyondBornOppenheimerEq_2} for the laboratory frame variables. With the preceding in mind, the Green's functions in the body-fixed frame, after tracing out the center-of-mass states, can be written as
\begin{equation} 
G'\left(\vec{y}t,\vec{y}'t'\right) = \frac{1}{i} \frac{\text{Tr}\left[e^{-\beta \hat{H}^{M}_{b}} \mathcal{T}\left\{ \hat{\phi}\left(\vec{y}t\right) \hat{\phi}^{\dagger}\left(\vec{y}'t'\right) \right\} \right]_{\Psi_{b}}}{\text{Tr}\left[e^{-\beta \hat{H}^{M}_{b}}\right]_{\Psi_{b}}},
\label{eq:BeyondBornOppenheimerEq_10}
\end{equation}
and
\begin{equation} 
D'_{\alpha_{\bar{n}}}\left(k_{\bar{n}}t_{\bar{n}}\right) = \frac{1}{i^{n-1}} \frac{\text{Tr}\left[e^{-\beta \hat{H}^{M}_{b}} \mathcal{T}\left\{ \hat{u}'_{\alpha_{\bar{n}}}\left(k_{\bar{n}}t_{\bar{n}}\right) \right\} \right]_{\Psi_{b}}}{\text{Tr}\left[e^{-\beta \hat{H}^{M}_{b}}\right]_{\Psi_{b}}}.
\label{eq:BeyondBornOppenheimerEq_11}
\end{equation}
These definitions are valid as such and there is no need to impose any additional conditions, like specific boundary conditions. The exact equations of motion for these Green's functions were obtained in Ref. \cite{Harkonen-ManybodyGreensFunctionTheoryOfElectronsAndNucleiBeyondTheBornOppenheimerApproximation-PhysRevB.101.235153-2020} and it was also justified that in the case of crystals the Hamiltonian can be approximated as 
\begin{equation} 
\hat{H}_{b} \approx \hat{T}'_{e} + \hat{T}'_{n} + \hat{V}'_{ee} + \hat{V}'_{en} + \hat{V}'_{nn} = \hat{H}'_{BO} + \hat{T}'_{n}.
\label{eq:BeyondBornOppenheimerEq_12}
\end{equation}
This result follows as the so-called mass polarization terms are proportional to the inverse of the nuclear mass of the system, which are negligible for crystals. Moreover, for Eq. \ref{eq:BeyondBornOppenheimerEq_12} to hold, the Coriolis and vibrational-rotational coupling terms have to be small, and this is the case in crystals when the Euler angles are connected to the nuclear variables through the Eckart condition \cite{Eckart-SomeStudiesConcerningRotatingAxesAndPolyatomicMolecules-PhysRev.47.552-1935,Harkonen-ManybodyGreensFunctionTheoryOfElectronsAndNucleiBeyondTheBornOppenheimerApproximation-PhysRevB.101.235153-2020}. The resulting functions $G'\left(\vec{y}t,\vec{y}'t'\right)$ and $D'_{\alpha_{\bar{n}}}\left(k_{\bar{n}}t_{\bar{n}}\right)$ resemble closely to those in Eqs. \ref{eq:BeyondBornOppenheimerEq_1} and \ref{eq:BeyondBornOppenheimerEq_2}. There are two differences, the Hamiltonian is now written in terms of the internal coordinates in the body-fixed frame instead of coordinates in the laboratory frame and here we traced out the center-of-mass states. The remaining states $\ket{\Psi_{b}}$ describe the internal motion, that is independent of the center-of-mass motion. In contrast, the original states $\ket{\Psi}$ are used as such in the Green's functions defined in the laboratory frame, Eqs. \ref{eq:BeyondBornOppenheimerEq_1} and \ref{eq:BeyondBornOppenheimerEq_2}.

\section{Many-Body Perturbation Theory in Exact Factorization}
\label{cha:ManyBodyPerturbationTheoryInExactFactorization}

\subsection{Exact expansion}
\label{cha:ExactExpansion}

The traces in Eqs. \ref{eq:BeyondBornOppenheimerEq_1}, \ref{eq:BeyondBornOppenheimerEq_2}, \ref{eq:BeyondBornOppenheimerEq_10} and \ref{eq:BeyondBornOppenheimerEq_11} are taken with respect to the states that belong to the full electron-nuclear space. We could take a rather straightforward approach and develop a many-body perturbation for these Green's functions by using the methods described in the literature \cite{Stefanucci-Leeuwen-many-body-book-2013}. However, this is probably not a useful thing to do from a practical point of view. Here we instead establish the exact factorization of the wave function and rewrite the Green's functions, like those given by Eqs. \ref{eq:BeyondBornOppenheimerEq_1}, \ref{eq:BeyondBornOppenheimerEq_2}, \ref{eq:BeyondBornOppenheimerEq_10} and \ref{eq:BeyondBornOppenheimerEq_11}, in terms of the exact factorized states. This allows us to derive beyond-BO approximations systematically, and we see that the approximations under which Eqs. \ref{eq:BeyondBornOppenheimerEq_1} and \ref{eq:BeyondBornOppenheimerEq_2} are useful can be deduced as special cases. We establish the exact factorization in the laboratory frame for the sake of simplicity, but the results are formally the same in the body-fixed frame for crystals when the Hamiltonian of Eq. \ref{eq:BeyondBornOppenheimerEq_12} is used. We provide in Appendix \ref{PerturbationExpansionInTheBodyFixedFrame} the body-fixed versions of the exact laboratory frame expansions derived in the following.

We start by writing the electronic Green's function in the contour formalism \cite{Stefanucci-Leeuwen-many-body-book-2013}, including Eq. \ref{eq:BeyondBornOppenheimerEq_1} as a special case, as follows
\begin{equation} 
G\left(\vec{y}z,\vec{y}'z'\right) = \frac{1}{i} \frac{\text{Tr}\left[\mathcal{T}_{c}\left\{ e^{-i \int_{c} d\bar{z} \hat{H}\left(\bar{z}\right)} \hat{\psi}\left(\vec{y}z\right) \hat{\psi}^{\dagger}\left(\vec{y}'z'\right) \right\} \right]_{\Psi}}{\text{Tr}\left[\mathcal{T}_{c}\left\{e^{-i \int_{c} d\bar{z} \hat{H}\left(\bar{z}\right)}\right\}\right]_{\Psi}},
\label{eq:ManyBodyPerturbationTheoryInExactFactorizationEq_1}
\end{equation}
and in a similar way for the nuclear Green's function given by Eq. \ref{eq:BeyondBornOppenheimerEq_2}. The contour formalism allows us to rewrite $G\left(\vec{y}t,\vec{y}'t'\right)$ such that the evolution operators and the exponential term in the ensemble average are included to a single exponent which is under the contour time-ordering $\mathcal{T}_{c}$ making the algebraic manipulations less complicated. We can extract the time-ordered function $G\left(\vec{y}t,\vec{y}'t'\right)$ and also other contour components from $G\left(\vec{y}z,\vec{y}'z'\right)$ by using the results described in the literature \cite{Stefanucci-Leeuwen-many-body-book-2013}. In the following, we will use the fact that under the contour time-ordering the operators either commute or anti-commute. We write the Hamiltonian as $\hat{H} = \hat{T}_{n} + \hat{H}_{BO}$ (see Sec. \ref{cha:HamiltonianAndBornOppenheimerApproximation}) and since the Hamiltonian is a sum of bosonic operators, these operators commute under the contour time-ordering, and we can write
\begin{equation} 
\mathcal{T}_{c}\left\{e^{-i \int_{c} d\bar{z} \hat{H}\left(\bar{z}\right)} \hat{o} \right\} = \mathcal{T}_{c}\left\{e^{-i \int_{c} d\bar{z} \hat{H}_{BO}\left(\bar{z}\right) } e^{-i \int_{c} d\bar{z} \hat{T}_{n}\left(\bar{z}\right) } \hat{o} \right\},
\label{eq:ManyBodyPerturbationTheoryInExactFactorizationEq_3}
\end{equation}
or alternatively
\begin{equation} 
\text{Tr}\left[ \mathcal{T}_{c}\left\{e^{-i \int_{c} d\bar{z} \hat{H}\left(\bar{z}\right)} \hat{o} \right\} \right]_{\Psi} = \text{Tr}\left[ \mathcal{T}_{c}\left\{ e^{-i \int_{c} d\bar{z} \hat{H}'_{n}\left(\bar{z}\right) } \hat{o} \right\} \right]_{\Psi}.
\label{eq:ManyBodyPerturbationTheoryInExactFactorizationEq_3_2}
\end{equation}
In Eq. \ref{eq:ManyBodyPerturbationTheoryInExactFactorizationEq_3_2}, we used Eqs. \ref{eq:PreliminariesEq_4} and \ref{eq:PreliminariesEq_5} and
\begin{equation} 
\hat{H}'_{n}\left(\bar{z}\right) \equiv \hat{T}_{n}\left(\bar{z}\right) + \hat{\epsilon}\left(\bar{z}\right) - \hat{U}_{en}\left(\bar{z}\right).
\label{eq:ManyBodyPerturbationTheoryInExactFactorizationEq_4}
\end{equation}
We now use Eq. \ref{eq:ManyBodyPerturbationTheoryInExactFactorizationEq_3} for the electronic Green's function together with the results obtained in Appendix \ref{ExactFactorizedPerturbationExpansion}. After the expansion of the nuclear kinetic energy, the numerator of Eq. \ref{eq:ManyBodyPerturbationTheoryInExactFactorizationEq_1} can be written as
\begin{eqnarray} 
&&\text{Tr}\left[ \mathcal{T}_{c}\left\{ e^{ -i \int_{c}d\bar{z} \hat{H}\left(\bar{z}\right) } \hat{\psi}\left(\vec{y}z\right) \hat{\psi}^{\dagger}\left(\vec{y}'z'\right) \right\} \right]_{\Psi} \nonumber \\
&&\ \ \ =  \sum_{m} \int d\vec{R} \sum^{\infty}_{s = 0} \frac{\left(-i\right)^{s}}{s!} \int_{c} d\bar{z}_{\bar{s}} \sum_{k_{\bar{s}}} \frac{ 1 }{2^{s} M_{k_{\bar{s}}}} \sum^{2}_{l_{\bar{s}} = 0} \binom{2}{l_{\bar{s}}}   \nonumber \\
&&\ \ \ \times \bra{\Phi^{\left(m\right)}_{\vec{R}}}\mathcal{T}_{c}\left\{ e^{ -i \int_{c}d\bar{z} \hat{H}_{BO}\left(\vec{R},\bar{z}\right) } \hat{\psi}\left(\vec{y}z\right) \hat{\psi}^{\dagger}\left(\vec{y}'z'\right) \right. \nonumber \\
&&\ \ \ \times \left. \vec{C}^{\left(2-l_{\bar{s}}\right)}_{mk_{\bar{s}}}\left(\vec{R},\bar{z}_{\bar{s}}\right) \vec{P}^{l_{\bar{s}}}_{k_{\bar{s}}}\left(\bar{z}_{\bar{s}}\right) \right\}\ket{\Phi^{\left(m\right)}_{\vec{R}}},
\label{eq:ManyBodyPerturbationTheoryInExactFactorizationEq_5}
\end{eqnarray}
and the denominator as
\begin{eqnarray} 
&&\text{Tr}\left[ \mathcal{T}_{c}\left\{ e^{ -i \int_{c}d\bar{z} \hat{H}\left(\bar{z}\right) } \right\} \right]_{\Psi} \nonumber \\
&&\ \ \ = \sum_{m} \int d\vec{R} \sum^{\infty}_{s = 0} \frac{\left(-i\right)^{s}}{s!} \int_{c} d\bar{z}_{\bar{s}} \sum_{k_{\bar{s}}} \frac{ 1 }{2^{s} M_{k_{\bar{s}}}} \sum^{2}_{l_{\bar{s}} = 0} \binom{2}{l_{\bar{s}}}  \nonumber \\
&&\ \ \ \times \bra{\Phi^{\left(m\right)}_{\vec{R}} } \mathcal{T}_{c}\left\{ e^{ -i \int_{c}d\bar{z} \hat{H}_{BO}\left(\vec{R},\bar{z}\right) } \right. \nonumber \\
&&\ \ \ \times \left. \vec{C}^{\left(2-l_{\bar{s}}\right)}_{mk_{\bar{s}}}\left(\vec{R},\bar{z}_{\bar{s}}\right) \vec{P}^{l_{\bar{s}}}_{k_{\bar{s}}}\left(\bar{z}_{\bar{s}}\right) \right\} \ket{\Phi^{\left(m\right)}_{\vec{R}}}.
\label{eq:ManyBodyPerturbationTheoryInExactFactorizationEq_6}
\end{eqnarray}
In Eqs. \ref{eq:ManyBodyPerturbationTheoryInExactFactorizationEq_5} and \ref{eq:ManyBodyPerturbationTheoryInExactFactorizationEq_6}
\begin{equation} 
\vec{C}^{\left(2-l_{\bar{s}}\right)}_{mk_{\bar{s}}}\left(\vec{R},\bar{z}_{\bar{s}}\right) \equiv \chi^{\ast}_{m}\left(\vec{R}\right) \vec{P}^{2-l_{\bar{s}}}_{k_{\bar{s}}}\left(\bar{z}_{\bar{s}}\right) \chi_{m}\left(\vec{R}\right).
\label{eq:ManyBodyPerturbationTheoryInExactFactorizationEq_7}
\end{equation}
The used notation for the nuclear momentum operators and so on is described in Appendix \ref{Notation}. The relations given by Eqs. \ref{eq:ManyBodyPerturbationTheoryInExactFactorizationEq_5} and \ref{eq:ManyBodyPerturbationTheoryInExactFactorizationEq_6} allow us to rewrite the exact electronic Green's function $G\left(\vec{y}z,\vec{y}'z'\right)$ in terms of the exact factorized states. We will consider the approximations derived from these relations in Sec. \ref{cha:ApproximationsAndConnectionToBornOppenheimerTheory}.

We obtain the perturbation expansion of the nuclear Green's functions in a similar way by using the methodology described in Appendix \ref{ExactFactorizedPerturbationExpansion}. Namely, by using Eq. \ref{eq:ManyBodyPerturbationTheoryInExactFactorizationEq_3_2} for the nuclear Green's function $D_{\alpha_{\bar{n}}}\left(k_{\bar{n}}z_{\bar{n}}\right)$ (the contour form, including Eq. \ref{eq:BeyondBornOppenheimerEq_2}), we expand the nuclear kinetic energy term and write the numerator as
\begin{eqnarray} 
&&\text{Tr}\left[ \mathcal{T}_{c} \left\{ e^{ -i \int_{c}d\bar{z} \hat{H}'_{n}\left(\bar{z}\right) } \hat{u}_{\alpha_{\bar{n}}}\left(k_{\bar{n}}z_{\bar{n}}\right) \right\} \right]_{\Psi} \nonumber \\
&&\ \ \ = \sum_{m} \int d\vec{r} \int d\vec{R} \sum^{\infty}_{s = 0} \frac{\left(-i\right)^{s}}{s!} \int_{c} d\bar{z}_{\bar{s}} \nonumber \\
&&\ \ \ \times \sum_{k_{\bar{s}}} \frac{ 1 }{2^{s} M_{k_{\bar{s}}}} \sum^{2}_{l_{\bar{s}} = 0} \binom{2}{l_{\bar{s}}} \Phi^{\left(m\right)\ast}_{\vec{R}}\left(\vec{r}\right) \chi^{\ast}_{m}\left(\vec{R}\right) \nonumber \\
&&\ \ \ \times \mathcal{T}_{c} \left\{ e^{ -i \int_{c}d\bar{z} \left[\epsilon_{m}\left(\bar{z}\right) - U^{\left(m\right)}_{en}\left(\bar{z}\right)\right] } u_{\alpha_{\bar{n}}}\left(k_{\bar{n}}z_{\bar{n}}\right) \right. \nonumber \\
&&\ \ \ \times \left. \left[\vec{P}^{l_{\bar{s}}}_{k_{\bar{s}}}\left(\bar{z}_{\bar{s}}\right) \Phi^{\left(m\right)}_{\vec{R}}\left(\vec{r}\right)\right] \left[ \vec{P}^{2-l_{\bar{s}}}_{k_{\bar{s}}}\left(\bar{z}_{\bar{s}}\right) \chi_{m}\left(\vec{R}\right)\right] \right\}, 
\label{eq:ManyBodyPerturbationTheoryInExactFactorizationEq_8}
\end{eqnarray}
and the denominator as
\begin{eqnarray} 
&&\text{Tr}\left[ \mathcal{T}_{c}\left\{ e^{ -i \int_{c}d\bar{z} \hat{H}'_{n}\left(\bar{z}\right) } \right\} \right]_{\Psi} \nonumber \\
&&\ \ \ = \sum_{m} \int d\vec{r} \int d\vec{R} \sum^{\infty}_{s = 0} \frac{\left(-i\right)^{s}}{s!} \int_{c} d\bar{z}_{\bar{s}} \nonumber \\
&&\ \ \ \times \sum_{k_{\bar{s}}} \frac{ 1 }{2^{s} M_{k_{\bar{s}}}} \sum^{2}_{l_{\bar{s}} = 0} \binom{2}{l_{\bar{s}}} \Phi^{\left(m\right)\ast}_{\vec{R}}\left(\vec{r}\right) \chi^{\ast}_{m}\left(\vec{R}\right) \nonumber \\
&&\ \ \ \times \mathcal{T}_{c}\left\{ e^{ -i \int_{c} d\bar{z} \left[\epsilon_{m}\left(\bar{z}\right) - U^{\left(m\right)}_{en}\left(\bar{z}\right)\right] } \left[\vec{P}^{l_{\bar{s}}}_{k_{\bar{s}}}\left(\bar{z}_{\bar{s}}\right) \Phi^{\left(m\right)}_{\vec{R}}\left(\vec{r}\right)\right] \right. \nonumber \\
&&\ \ \ \times \left. \left[ \vec{P}^{2-l_{\bar{s}}}_{k_{\bar{s}}}\left(\bar{z}_{\bar{s}}\right) \chi_{m}\left(\vec{R}\right)\right] \right\}.
\label{eq:ManyBodyPerturbationTheoryInExactFactorizationEq_9}
\end{eqnarray}
With Eqs. \ref{eq:ManyBodyPerturbationTheoryInExactFactorizationEq_8} and \ref{eq:ManyBodyPerturbationTheoryInExactFactorizationEq_9}, we can rewrite the exact nuclear Green's function $D_{\alpha_{\bar{n}}}\left(k_{\bar{n}}z_{\bar{n}}\right)$ in terms of the exact factorized states and we consider the approximations in Sec. \ref{cha:ApproximationsAndConnectionToBornOppenheimerTheory}. The operator $U^{\left(m\right)}_{en}$ contains differential operators on the nuclear variables and we have to take this into account when ordering the terms like $\vec{P}^{l_{\bar{s}}}_{k_{\bar{s}}}\left(\bar{z}_{\bar{s}}\right) \Phi^{\left(m\right)}_{\vec{R}}\left(\vec{r}\right)$ and $\vec{P}^{2-l_{\bar{s}}}_{k_{\bar{s}}}\left(\bar{z}_{\bar{s}}\right) \chi_{m}\left(\vec{R}\right)$ within the time ordering. That is, we cannot change the order of the terms $e^{ -i \int_{c} d\bar{z} \left[\epsilon_{m}\left(\bar{z}\right) - U^{\left(m\right)}_{en}\left(\bar{z}\right)\right] }$ and $\vec{P}^{l_{\bar{s}}}_{k_{\bar{s}}}\left(\bar{z}_{\bar{s}}\right) \Phi^{\left(m\right)}_{\vec{R}}\left(\vec{r}\right)$, say, without indicating this in our notation. The correct ordering is important as in our notation the differential operators act only on the right.

To understand the physical picture what is being established here, consider Eq. \ref{eq:ManyBodyPerturbationTheoryInExactFactorizationEq_1}. The Green's function $G\left(\vec{y}'t',\vec{y}'z'\right)$ is a time-dependent ensemble average of time-ordering of the operator $\hat{\psi}\left(\vec{y}z\right) \hat{\psi}^{\dagger}\left(\vec{y}'z'\right)$, one of the special cases of which is the time-dependent electron density $n_{e}\left(\vec{y}t\right) = -i G\left(\vec{y}t,\vec{y}t^{+}\right)$. The ensemble average here contains two different expectation values: the thermal averaging and the quantum mechanical expectation with respect to the states $\ket{\Psi}$. Both of these expectations are taken with respect to the exact Hamiltonian $\hat{H}$ defined by Eq. \ref{eq:PreliminariesEq_1}. When we use Eq. \ref{eq:ManyBodyPerturbationTheoryInExactFactorizationEq_3} and expand in the nuclear kinetic energy, we obtain to the lowest order, the thermal averaging and the time-evolution with respect to the BO Hamiltonian. By taking into account the higher order nuclear kinetic energy expansion terms, we include corrections to the thermal averaging and the time evolution, induced by the nuclear motion. The higher order corrections produce more complicated expected values to be calculated, but all operators will be in the Heisenberg picture defined with respect to the BO Hamiltonian, with respect to which the thermal averaging is also established. At this point, we still compute the resulting quantum mechanical expectations with respect to the states $\ket{\Psi}$. The gain here is that the time-evolution and the thermal averaging is with respect to the $\hat{H}_{BO}$. The price we pay is the more complicated Green's functions to compute, involving the products of nuclear momenta and electronic field operators. After the exact factorization of $\ket{\Psi}$, we write the quantum mechanical expectation with respect to the conditional $\Phi_{\vec{R}}\left(\vec{r}\right)$ and marginal functions $\chi\left(\vec{R}\right)$. Consequently, various terms originating from the nuclear momentum operators, like $\vec{C}^{\left(2-l_{\bar{s}}\right)}_{mk_{\bar{s}}}\left(\vec{R},\bar{z}_{\bar{s}}\right)$, will appear. The gain here is that we have some grasp how to compute the expected values as these are now with respect to the conditional electronic states and states in the nuclear space only. The same line of thought also applies to the nuclear Green's functions.

In summary, the lowest order terms in the nuclear kinetic energy expansion will usually be the most significant ones. In this case, the remaining electronic Green's functions are defined with respect to $\hat{H}_{BO}$ and contain one or few nuclear momentum operators. These quantum mechanical expectations are essentially written in terms of exact factorized states, which at least in these lowest order cases, make computations of these quantities more accessible, when compared with the complexity of the original Green's functions of Eqs. \ref{eq:BeyondBornOppenheimerEq_1} and \ref{eq:BeyondBornOppenheimerEq_2}.

\subsection{Approximations and connection to Born-Oppenheimer theory}
\label{cha:ApproximationsAndConnectionToBornOppenheimerTheory}

We can find the BO Green's functions as special cases of our exact results so far. We first look the electronic Green's function $G\left(\vec{r}z,\vec{r}'z'\right)$. To lowest order, Eqs. \ref{eq:ManyBodyPerturbationTheoryInExactFactorizationEq_5} and \ref{eq:ManyBodyPerturbationTheoryInExactFactorizationEq_6} become
\begin{eqnarray} 
&&\text{Tr}\left[ \mathcal{T}_{c}\left\{ e^{ -i \int_{c}d\bar{z} \hat{H}\left(\bar{z}\right) } \hat{\psi}\left(\vec{y}z\right) \hat{\psi}^{\dagger}\left(\vec{y}'z'\right) \right\} \right]_{\Psi} \nonumber \\
&&\ \ \ \approx  \sum_{m} \int d\vec{R} \left|\chi_{m}\left(\vec{R}\right)\right|^{2}  \nonumber \\
&&\ \ \ \times \braket{\Phi^{\left(m\right)}_{\vec{R}} | \mathcal{T}_{c}\left\{ e^{ -i \int_{c}d\bar{z} \hat{H}_{BO}\left(\vec{R},\bar{z}\right) } \hat{\psi}\left(\vec{y}z\right) \hat{\psi}^{\dagger}\left(\vec{y}'z'\right) \right\} | \Phi^{\left(m\right)}_{\vec{R}}}, \nonumber \\
\label{eq:ApproximationsAndConnectionToBornOppenheimerTheoryEq_1}
\end{eqnarray}
and
\begin{eqnarray} 
&&\text{Tr}\left[ \mathcal{T}_{c}\left\{ e^{ -i \int_{c}d\bar{z} \hat{H}\left(\bar{z}\right) } \right\} \right]_{\Psi} \nonumber \\
&&\ \ \ \approx \sum_{m} \int d\vec{R} \left|\chi_{m}\left(\vec{R}\right)\right|^{2} \nonumber \\
&&\ \ \ \times \braket{\Phi^{\left(m\right)}_{\vec{R}} | \mathcal{T}_{c}\left\{ e^{ -i \int_{c}d\bar{z} \hat{H}_{BO}\left(\vec{R},\bar{z}\right) } \right\} | \Phi^{\left(m\right)}_{\vec{R}}}.
\label{eq:ApproximationsAndConnectionToBornOppenheimerTheoryEq_2}
\end{eqnarray}
By combining Eqs. \ref{eq:ApproximationsAndConnectionToBornOppenheimerTheoryEq_1} and \ref{eq:ApproximationsAndConnectionToBornOppenheimerTheoryEq_2} we can write the lowest order approximation for $G\left(\vec{r}z,\vec{r}'z'\right)$ of Eq. \ref{eq:WithinBornOppenheimerEq_1}. For simplicity, we take the zero temperature limit of Eqs. \ref{eq:ApproximationsAndConnectionToBornOppenheimerTheoryEq_1} and \ref{eq:ApproximationsAndConnectionToBornOppenheimerTheoryEq_2} and extract the time-ordered component. In this case, the approximate form of $G\left(\vec{r}t,\vec{r}'t'\right)$ can be written as
\begin{equation} 
G\left(\vec{y}t,\vec{y}'t'\right) \approx \int d\vec{R} \left|\chi\left(\vec{R}\right)\right|^{2} G^{BO}_{\vec{R}}\left(\vec{y}t,\vec{y}'t'\right),
\label{eq:ApproximationsAndConnectionToBornOppenheimerTheoryEq_3}
\end{equation}
where we denote
\begin{equation} 
G^{BO}_{\vec{R}}\left(\vec{y}t,\vec{y}'t'\right) = -i \braket{\Phi_{\vec{R}}| \mathcal{T}\left\{ \hat{\psi}\left(\vec{y}t\right) \hat{\psi}^{\dagger}\left(\vec{y}'t'\right) \right\}|\Phi_{\vec{R}}}.
\label{eq:ApproximationsAndConnectionToBornOppenheimerTheoryEq_4}
\end{equation}
In Eq. \eqref{eq:ApproximationsAndConnectionToBornOppenheimerTheoryEq_4}, the field operators are in the Heisenberg picture defined with respect to the Hamiltonian $\hat{H}_{BO}$. The Green's function of Eq. \ref{eq:ApproximationsAndConnectionToBornOppenheimerTheoryEq_4} is the zero temperature limit of Eq. \ref{eq:WithinBornOppenheimerEq_1}. The result given by Eq. \ref{eq:ApproximationsAndConnectionToBornOppenheimerTheoryEq_3} states that the exact electronic Green's function can be approximated as an expected value of the BO electronic Green's function $G^{BO}_{\vec{R}}\left(\vec{y}t,\vec{y}'t'\right)$ relative to the nuclear density $\left|\chi\left(\vec{R}\right)\right|^{2}$, the random variables being the nuclear coordinates $\vec{R}$. If the nuclei $\vec{R}$ are localized to their equilibrium positions $\vec{x}$, then $\left|\chi\left(\vec{R}\right)\right|^{2} = \delta\left(\vec{R}-\vec{x}\right)$ and Eq. \ref{eq:ApproximationsAndConnectionToBornOppenheimerTheoryEq_3} becomes $G\left(\vec{y}t,\vec{y}'t'\right) \approx G^{BO}_{\vec{x}}\left(\vec{y}t,\vec{y}'t'\right)$ which is the BO electronic Green's function. This is the approximation also made for $G\left(\vec{r}t,\vec{r}'t'\right)$ in the practical applications of the Hedin-Baym equations \cite{Giustino-ElectronPhononInteractFromFirstPrinc-RevModPhys.89.015003-2017}. The same holds for the finite temperature case when we approximate $\left|\chi^{\left(m\right)}\left(\vec{R}\right)\right|^{2} \approx \delta\left(\vec{R}-\vec{x}\right)$ in Eqs. \ref{eq:ApproximationsAndConnectionToBornOppenheimerTheoryEq_1} and \ref{eq:ApproximationsAndConnectionToBornOppenheimerTheoryEq_2} which leads to $G^{BO}_{\vec{R}}\left(\vec{y}t,\vec{y}'t'\right)$ of Eq. \ref{eq:WithinBornOppenheimerEq_1}. The more quantum mechanically the nuclei behave, the more uncertain their position is and the wider the distribution $\left|\chi\left(\vec{R}\right)\right|^{2}$. The wider the distribution the larger the discrepancy, in general, between the functions $G^{BO}_{\vec{R}}\left(\vec{y}t,\vec{y}'t'\right)$ and $G\left(\vec{y}t,\vec{y}'t'\right)$. For sufficiently broad nuclear distributions, we expect the electronic quantities, like the electron density, to change from the BO values making the approximation of Eq. \ref{eq:ApproximationsAndConnectionToBornOppenheimerTheoryEq_3} important when assessing the physical properties. Since $\left|\chi\left(\vec{R}\right)\right|^{2}$ is independent of time, the equations of motion for $G\left(\vec{y}t,\vec{y}'t'\right)$ given by Eq. \ref{eq:ApproximationsAndConnectionToBornOppenheimerTheoryEq_3} will be the well-known equations of motion for $G^{BO}_{\vec{R}}\left(\vec{y}t,\vec{y}'t'\right)$, weighted by the nuclear density.

\begin{figure}
\includegraphics[]{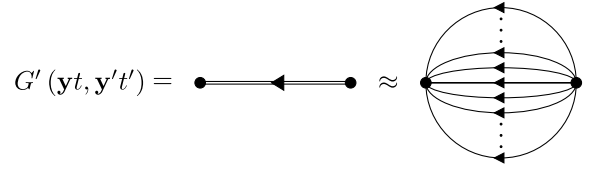}
\caption{The lowest order diagram for the electronic Green's function $G'\left(\vec{y}t,\vec{y}'t'\right)$ corresponding to Eq. \ref{eq:ApproximationsAndConnectionToBornOppenheimerTheoryEq_5}. On the right hand side, each single line denotes the BO electronic Green's function $G'^{BO}_{\vec{R}'}\left(\vec{y}t,\vec{y}'t'\right)$ for a particular value of $\vec{R}'$.} 
\label{fig:GreensFunctionDiagram}
\end{figure}

It is important to notice, however, that Eq. \ref{eq:ApproximationsAndConnectionToBornOppenheimerTheoryEq_3} as such will not give any physically relevant results, as justified in Sec. \ref{cha:Symmetry}. This laboratory frame form is only useful if we approximate $\left|\chi\left(\vec{R}\right)\right|^{2} \approx \delta\left(\vec{R}-\vec{x}\right)$, as is essentially done in the literature when the Hedin-Baym equations are applied in the actual computations. Therefore, the laboratory frame formulation of the electronic part is useful only strictly in the BO approximation. The good news is that we obtain formally the same result in the body-fixed frame when the Hamiltonian is given by Eq. \ref{eq:BeyondBornOppenheimerEq_12}. That is, we can use Eqs. \ref{eq:PerturbationExpansionInTheBodyFixedFrameEq_1} and \ref{eq:PerturbationExpansionInTheBodyFixedFrameEq_2} of Appendix \ref{PerturbationExpansionInTheBodyFixedFrame} to the lowest order and approximate $G'\left(\vec{y}t,\vec{y}'t'\right)$ at the zero temperature limit as
\begin{equation} 
G'\left(\vec{y}t,\vec{y}'t'\right) \approx \int d\vec{R}' \left|\chi\left(\vec{R}'\right)\right|^{2} G'^{BO}_{\vec{R}'}\left(\vec{y}t,\vec{y}'t'\right),
\label{eq:ApproximationsAndConnectionToBornOppenheimerTheoryEq_5}
\end{equation}
where $G'^{BO}_{\vec{R}'}\left(\vec{y}t,\vec{y}'t'\right)$ is defined as in Eq. \ref{eq:ApproximationsAndConnectionToBornOppenheimerTheoryEq_4} but with respect to the body-fixed BO Hamiltonian $\hat{H}'_{BO}$ which is written in terms of variables in the body-fixed frame, see Ref. \cite{Harkonen-ManybodyGreensFunctionTheoryOfElectronsAndNucleiBeyondTheBornOppenheimerApproximation-PhysRevB.101.235153-2020}. To emphasize the difference of $G'\left(\vec{y}t,\vec{y}'t'\right)$ and $G'^{BO}_{\vec{R}'}\left(\vec{y}t,\vec{y}'t'\right)$, the diagram corresponding to Eq. \ref{eq:ApproximationsAndConnectionToBornOppenheimerTheoryEq_5} is depicted in Fig. \ref{fig:GreensFunctionDiagram}. We need the BO Green's function $G'^{BO}_{\vec{R}'}\left(\vec{y}t,\vec{y}'t'\right)$ for all values of $\vec{R}'$ in order to compute $G'\left(\vec{y}t,\vec{y}'t'\right)$. Each of the separate single line in Fig. \ref{fig:GreensFunctionDiagram} denote a Green's function $G'^{BO}_{\vec{R}'}\left(\vec{y}t,\vec{y}'t'\right)$ for a particular value of $\vec{R}'$. In the BO approximation, we only need to consider one particular line on the right-hand side diagram of \ref{fig:GreensFunctionDiagram}.

To obtain an approximation to the nuclear Green's function of Eq. \ref{eq:BeyondBornOppenheimerEq_2} by using Eqs. \ref{eq:ManyBodyPerturbationTheoryInExactFactorizationEq_8} and \ref{eq:ManyBodyPerturbationTheoryInExactFactorizationEq_9}, we consider the case $l_{\bar{s}} = 0$ and neglect all those terms where $U_{en}$ acts on $\Phi_{\vec{R}}$, in this case
\begin{equation} 
D_{\alpha_{\bar{n}}}\left(k_{\bar{n}}z_{\bar{n}}\right) \approx \frac{1}{i^{n-1}} \frac{ \text{Tr}\left[ \mathcal{T}_{c} \left\{ e^{ -i \int_{c}d\bar{z} \hat{H}'_{n}\left(\bar{z}\right) } \hat{u}_{\alpha_{\bar{n}}}\left(k_{\bar{n}}z_{\bar{n}}\right) \right\} \right]_{\chi} }{ \text{Tr}\left[ \mathcal{T}_{c}\left\{ e^{ -i \int_{c}d\bar{z} \hat{H}'_{n}\left(\bar{z}\right) } \right\} \right]_{\chi} }.
\label{eq:ApproximationsAndConnectionToBornOppenheimerTheoryEq_6}
\end{equation}
If we further approximate $\hat{U}_{en} \approx 0$ and $\hat{\epsilon} \approx \hat{\epsilon}_{BO}$, the Hamiltonian in Eq. \ref{eq:ApproximationsAndConnectionToBornOppenheimerTheoryEq_6}, defined by Eq. \ref{eq:ManyBodyPerturbationTheoryInExactFactorizationEq_4}, becomes $\hat{H}'_{n} \approx \hat{T}_{n} + \hat{\epsilon}_{BO} = \hat{H}_{ph}$, which is the Hamiltonian of Eq. \ref{eq:PreliminariesEq_8}. In this case, Eq. \ref{eq:ApproximationsAndConnectionToBornOppenheimerTheoryEq_6} becomes the BO nuclear Green's function given by Eq. \ref{eq:WithinBornOppenheimerEq_3} when the time-ordered component is extracted and we again obtained the BO theory as a special case of our exact approach. We have already shown that the laboratory frame formulation is valid only in the strict BO approximation. Therefore, in all implementations of the beyond-BO Green's function theory we have to use the body-fixed formulation. In the case of nuclear Green's functions $D'_{\alpha_{\bar{n}}}\left(k_{\bar{n}}z_{\bar{n}}\right)$, the lowest order approximation corresponding to Eq. \ref{eq:ApproximationsAndConnectionToBornOppenheimerTheoryEq_6}, and the higher order approximations, can be obtained from the exact relations given by Eqs. \ref{eq:PerturbationExpansionInTheBodyFixedFrameEq_4} and \ref{eq:PerturbationExpansionInTheBodyFixedFrameEq_5}.

\section{Conclusions}
\label{cha:Conclusions}

In this work, we combined the many-body Green's function and exact factorization approaches to describe non-relativistic quantum mechanical many-body systems of electrons and nuclei. We discussed the limitations of the laboratory frame formulation of the many-body Green's function theory and derived approximations. It was shown that in the laboratory frame formulation, the electronic Green's function is a useful quantity only by assuming the strict BO approximation, which is already done in the existing implementations. In other words, already the lowest order term in the expansion contains the nuclear density which renders the electronic Green's function useless in the laboratory frame formulation, unless the nuclear density itself is approximated by the delta function in the electronic problem.

We derived an exact expansion of the electronic and nuclear Green's functions in the nuclear kinetic energy by using the exact factorization approach. This allows systematic approximations for these functions beyond the BO approximation. The states with respect to which the expected values are taken are the exact factorized states instead of the general many-body states in the full electron-nuclear space. We showed how the BO many-body Green's functions follow as special cases of our exact approach. The lowest order approximation to the exact electronic Green's function was found to have a rather clear cut interpretation as an expectation value of the electronic BO Green's function with respect to the nuclear density. The simplest approximation to the nuclear Green's function adds more potential terms to the BO nuclear Hamiltonian, only changing the potential energy felt by the nuclei.

The steps taken here will make the implementations of the general Green's function theory derived in Ref. \cite{Harkonen-ManybodyGreensFunctionTheoryOfElectronsAndNucleiBeyondTheBornOppenheimerApproximation-PhysRevB.101.235153-2020} more accessible. We also took the first steps towards the many-body Green's function theory of exact factorization which has not been attempted before. As the results of this work imply, the beyond BO Green's function theory cannot be formulated in the laboratory frame and the body-fixed approaches are needed for this task. The progress made in this work takes us closer to the implementation of the many-body Green's function theory of electrons and nuclei beyond the BO approximation. We expect that these methods will become an important part of the tool box used in the description of systems, like molecules and solids, whenever the validity of the BO approximation is compromised.

\appendix

\section{Notation}
\label{Notation}

The following shorthand notations are used in this work
\begin{eqnarray} 
\vec{r} &\equiv& \vec{r}_{1}, \ldots, \vec{r}_{N_{e}}, \nonumber \\
\vec{R} &\equiv& \vec{R}_{1}, \ldots, \vec{R}_{N_{n}}, \nonumber \\
\vec{x} &\equiv& \vec{x}_{1}, \ldots, \vec{x}_{N_{n}}, \nonumber \\
\vec{u} &\equiv& \vec{u}_{1}, \ldots, \vec{u}_{N_{n}},
\label{eq:NotationEq_1}
\end{eqnarray}
and moreover
\begin{eqnarray} 
\alpha_{\bar{n}} &\equiv& \alpha_{1} \cdots \alpha_{n}, \nonumber \\
k_{\bar{n}}t_{\bar{n}} &\equiv& k_{1}t_{1},\ldots, k_{n}t_{n}, \nonumber \\
\hat{u}_{\alpha_{\bar{n}}}\left(k_{\bar{n}}t_{\bar{n}}\right) &\equiv& \hat{u}_{\alpha_{1}}\left(k_{1}t_{1}\right) \cdots \hat{u}_{\alpha_{n}}\left(k_{n}t_{n}\right),
\label{eq:NotationEq_2}
\end{eqnarray}
where $k$ labels the nucleus and $\alpha_{j}$ the Cartesian component. We use the following notation for the nuclear momentum operators (used for instance in Eqs. \ref{eq:ManyBodyPerturbationTheoryInExactFactorizationEq_5}, \ref{eq:ManyBodyPerturbationTheoryInExactFactorizationEq_6} and \ref{eq:ManyBodyPerturbationTheoryInExactFactorizationEq_7})
\begin{eqnarray} 
\vec{P}^{l_{\bar{s}}}_{k_{\bar{s}}}\left(z_{\bar{s}}\right) &\equiv& \vec{P}^{l_{1}}_{k_{1}}\left(z_{1}\right) \cdots \vec{P}^{l_{s}}_{k_{s}}\left(z_{s}\right), \nonumber \\
\vec{P}^{2-l_{\bar{s}}}_{k_{\bar{s}}}\left(z_{\bar{s}}\right) &\equiv& \vec{P}^{2-l_{1}}_{k_{1}}\left(z_{1}\right) \cdots \vec{P}^{2-l_{s}}_{k_{s}}\left(z_{s}\right),
\label{eq:NotationEq_3}
\end{eqnarray}
where $\vec{P}_{k_{j}} = -i \nabla_{\vec{R}_{k_{j}}}$ and $s, l_{j}$ some integers greater or equal to zero. For sums, integrals, the products of binomial coefficients and so on, we use the following shorthand notations
\begin{eqnarray} 
\int_{c} d\bar{z}_{\bar{s}} &\equiv& \int_{c} d\bar{z}_{1} \cdots \int_{c} d\bar{z}_{s}, \nonumber \\
\sum_{k_{\bar{s}}} &\equiv& \sum_{k_{1}, \ldots, k_{s}}, \quad \sum^{2}_{l_{\bar{s}} = 0} \equiv \sum^{2}_{l_{1}, \ldots, l_{s} = 0}, \nonumber \\
M_{k_{\bar{s}}} &\equiv& M_{k_{1}} \cdots M_{k_{s}}, \quad \binom{2}{l_{\bar{s}}} \equiv \binom{2}{l_{1}} \cdots \binom{2}{l_{s}}. 
\label{eq:NotationEq_4}
\end{eqnarray}
In Eq. \ref{eq:NotationEq_4}, $\binom{2}{l_{j}}$ is the binomial coefficient.

\section{Exact factorized perturbation expansion}
\label{ExactFactorizedPerturbationExpansion}

Here we derive some results needed in the perturbation expansions given in Sec. \ref{cha:ManyBodyPerturbationTheoryInExactFactorization} and Appendix \ref{PerturbationExpansionInTheBodyFixedFrame}. We write for a general state in the full electron-nuclei space
\begin{equation} 
\ket{\Psi} = \int d\vec{r} \int d\vec{R} \Phi_{\vec{R}}\left(\vec{r}\right) \chi\left(\vec{R}\right) \ket{\vec{r}} \otimes \ket{\vec{R}},
\label{eq:ExactFactorizedPerturbationExpansionEq_1}
\end{equation}
where we established the exact factorization given by Eq. \ref{eq:PreliminariesEq_3}. The momentum operator of the $k$th nucleus can be written as
\begin{equation} 
\hat{\vec{P}}_{k} = \int d\vec{R} \vec{P}_{k}  \ket{\vec{R}} \bra{\vec{R}}, \quad \vec{P}_{k} = -i \nabla_{\vec{R}_{k}}.
\label{eq:ExactFactorizedPerturbationExpansionEq_2}
\end{equation}
We act with $\hat{\vec{P}}^{2n}_{k}$ on $\ket{\Psi}$ given by Eq. \ref{eq:ExactFactorizedPerturbationExpansionEq_1} and after using the product rule for the derivatives \cite{Hardy-CombinatoricsOfPartialDerivatives-2006}
\begin{eqnarray} 
\hat{\vec{P}}^{2n}_{k} \ket{\Psi} &=& \int d\vec{r} \int d\vec{R} \sum^{2n}_{s = 0} \binom{2n}{s} \left[\vec{P}^{2n-s}_{k} \chi\left(\vec{R}\right)\right] \nonumber \\
&&\times \left[\vec{P}^{s}_{k} \Phi_{\vec{R}}\left(\vec{r}\right)\right] \ket{\vec{r}} \otimes \ket{\vec{R}}.
\label{eq:ExactFactorizedPerturbationExpansionEq_3}
\end{eqnarray}
The nuclear kinetic energy is of the form $\hat{T}_{n} = \sum_{k} \hat{\vec{P}}^{2}_{k} / 2 M_{k}$ and thus, by making use of Eq. \eqref{eq:ExactFactorizedPerturbationExpansionEq_3}, we find that
\begin{eqnarray} 
&&\hat{T}_{n}\left(z_{\bar{s}}\right) \ket{\Psi} \nonumber \\
&&\ \ \ = \int d\vec{r} \int d\vec{R} \sum_{k_{\bar{s}}} \frac{ 1 }{2^{s} M_{k_{\bar{s}}}} \sum^{2}_{l_{\bar{s}} = 0} \binom{2}{l_{\bar{s}}} \left[\vec{P}^{l_{\bar{s}}}_{k_{\bar{s}}}\left(z_{\bar{s}}\right) \Phi_{\vec{R}}\left(\vec{r}\right)\right] \nonumber \\
&&\ \ \ \times \left[ \vec{P}^{2-l_{\bar{s}}}_{k_{\bar{s}}}\left(z_{\bar{s}}\right) \chi\left(\vec{R}\right)\right] \ket{\vec{r}} \otimes \ket{\vec{R}},
\label{eq:ExactFactorizedPerturbationExpansionEq_4}
\end{eqnarray}
or
\begin{eqnarray} 
&&T_{n}\left(z_{\bar{s}}\right) \Phi_{\vec{R}}\left(\vec{r}\right) \chi\left(\vec{R}\right) \nonumber \\
&&\ \ \ = \sum_{k_{\bar{s}}} \frac{ 1 }{2^{s} M_{k_{\bar{s}}}} \sum^{2}_{l_{\bar{s}} = 0} \binom{2}{l_{\bar{s}}} \left[\vec{P}^{l_{\bar{s}}}_{k_{\bar{s}}}\left(z_{\bar{s}}\right) \Phi_{\vec{R}}\left(\vec{r}\right)\right] \nonumber \\
&&\ \ \ \times \left[ \vec{P}^{2-l_{\bar{s}}}_{k_{\bar{s}}}\left(z_{\bar{s}}\right) \chi\left(\vec{R}\right)\right].
\label{eq:ExactFactorizedPerturbationExpansionEq_5}
\end{eqnarray}
Here we used $T_{n}\left(\bar{z}_{\bar{s}}\right) = T_{n}\left(\bar{z}_{1}\right) \cdots T_{n}\left(\bar{z}_{s}\right)$ and the notation defined by Eqs. \ref{eq:NotationEq_3} and \ref{eq:NotationEq_4}. To expand the Green's functions, we have to consider, for instance, quantities of the following form
\begin{eqnarray} 
&&\text{Tr}\left[ \mathcal{T}_{c}\left\{ e^{ -i \int_{c}d\bar{z} \hat{H}_{BO}\left(\bar{z}\right) } e^{ -i \int_{c}d\bar{z} \hat{T}_{n}\left(\bar{z}\right) } \hat{o} \right\} \right]_{\Psi} \nonumber \\
&&\ \ \ = \sum_{m} \int d\vec{r} \int d\vec{R} \sum^{\infty}_{s = 0} \frac{\left(-i\right)^{s}}{s!} \int_{c} d\bar{z}_{\bar{s}} \Phi^{\left(m\right)\ast}_{\vec{R}}\left(\vec{r}\right) \chi^{\ast}_{m}\left(\vec{R}\right) \nonumber \\
&&\ \ \ \times \mathcal{T}_{c}\left\{ e^{ -i \int_{c}d\bar{z} H_{BO}\left(\bar{z}\right) } o\left(\vec{r},\vec{R}\right) T_{n}\left(\bar{z}_{\bar{s}}\right) \right\} \nonumber \\
&&\ \ \ \times \Phi^{\left(m\right)}_{\vec{R}}\left(\vec{r}\right) \chi_{m}\left(\vec{R}\right).
\label{eq:ExactFactorizedPerturbationExpansionEq_6}
\end{eqnarray}
where we expanded $e^{ -i \int_{c}d\bar{z} \hat{T}_{n}\left(\bar{z}\right) }$ and established the exact factorization given by Eq. \ref{eq:ExactFactorizedPerturbationExpansionEq_1}. In Eq. \ref{eq:ExactFactorizedPerturbationExpansionEq_6}, $\hat{o}$ is some operator acting in the electronic, nuclear or electron-nuclear space, including the identity. We use Eq. \ref{eq:ExactFactorizedPerturbationExpansionEq_6} combined with Eq. \ref{eq:ExactFactorizedPerturbationExpansionEq_5} to derive the perturbation expansions for the electronic Green's functions in Sec. \ref{cha:ManyBodyPerturbationTheoryInExactFactorization} and Appendix \ref{PerturbationExpansionInTheBodyFixedFrame}. For the nuclear Green's function the expansion in nuclear kinetic energy can be obtained in a similar way.

\section{Perturbation expansion in the body-fixed frame}
\label{PerturbationExpansionInTheBodyFixedFrame}

Here we derive the exact expansions of the Green's functions given by Eqs. \ref{eq:BeyondBornOppenheimerEq_10} and \ref{eq:BeyondBornOppenheimerEq_11} in the body-fixed frame by using the Hamiltonian of Eq. \ref{eq:BeyondBornOppenheimerEq_12}. Here, the Hamiltonian $\hat{H}_{b}$ is formally of the same form as the Hamiltonian $\hat{H}$ of Eq. \ref{eq:PreliminariesEq_1}. Thus, the exact factorization of the Green's functions in the body-fixed frame can be established as in Sec. \ref{cha:ExactExpansion} with a minor adjustment in notation. For the expansion of the electronic Green's function given by Eq. \ref{eq:BeyondBornOppenheimerEq_10}, we write for the numerator (compare to Eq. \ref{eq:ManyBodyPerturbationTheoryInExactFactorizationEq_5})
\begin{eqnarray} 
&&\text{Tr}\left[ \mathcal{T}_{c}\left\{ e^{ -i \int_{c}d\bar{z} \hat{H}_{b}\left(\bar{z}\right) } \hat{\phi}\left(\vec{y}z\right) \hat{\phi}^{\dagger}\left(\vec{y}'z'\right) \right\} \right]_{\Psi_{b}} \nonumber \\
&&\ \ \ =  \sum_{m} \int d\vec{R}' \sum^{\infty}_{s = 0} \frac{\left(-i\right)^{s}}{s!} \int_{c} d\bar{z}_{\bar{s}} \sum_{k_{\bar{s}}} \frac{ 1 }{2^{s} M_{k_{\bar{s}}}} \sum^{2}_{l_{\bar{s}} = 0} \binom{2}{l_{\bar{s}}}   \nonumber \\
&&\ \ \ \times \bra{\Phi^{\left(m\right)}_{\vec{R}'}}\mathcal{T}_{c}\left\{ e^{ -i \int_{c}d\bar{z} \hat{H}'_{BO}\left(\vec{R}',\bar{z}\right) } \hat{\phi}\left(\vec{y}z\right) \hat{\phi}^{\dagger}\left(\vec{y}'z'\right) \right. \nonumber \\
&&\ \ \ \times \left. \vec{C}'^{\left(2-l_{\bar{s}}\right)}_{mk_{\bar{s}}}\left(\vec{R}',\bar{z}_{\bar{s}}\right) \vec{P}'^{l_{\bar{s}}}_{k_{\bar{s}}}\left(\bar{z}_{\bar{s}}\right) \right\}\ket{\Phi^{\left(m\right)}_{\vec{R}'}},
\label{eq:PerturbationExpansionInTheBodyFixedFrameEq_1}
\end{eqnarray}
and for the denominator (compare to Eq. \ref{eq:ManyBodyPerturbationTheoryInExactFactorizationEq_6})
\begin{eqnarray} 
&&\text{Tr}\left[ \mathcal{T}_{c}\left\{ e^{ -i \int_{c}d\bar{z} \hat{H}_{b}\left(\bar{z}\right) } \right\} \right]_{\Psi_{b}} \nonumber \\
&&\ \ \ = \sum_{m} \int d\vec{R}' \sum^{\infty}_{s = 0} \frac{\left(-i\right)^{s}}{s!} \int_{c} d\bar{z}_{\bar{s}} \sum_{k_{\bar{s}}} \frac{ 1 }{2^{s} M_{k_{\bar{s}}}} \sum^{2}_{l_{\bar{s}} = 0} \binom{2}{l_{\bar{s}}}  \nonumber \\
&&\ \ \ \times \bra{\Phi^{\left(m\right)}_{\vec{R}'} } \mathcal{T}_{c}\left\{ e^{ -i \int_{c}d\bar{z} \hat{H}'_{BO}\left(\vec{R}',\bar{z}\right) } \right. \nonumber \\
&&\ \ \ \times \left. \vec{C}'^{\left(2-l_{\bar{s}}\right)}_{mk_{\bar{s}}}\left(\vec{R}',\bar{z}_{\bar{s}}\right) \vec{P}'^{l_{\bar{s}}}_{k_{\bar{s}}}\left(\bar{z}_{\bar{s}}\right) \right\} \ket{\Phi^{\left(m\right)}_{\vec{R}'}},
\label{eq:PerturbationExpansionInTheBodyFixedFrameEq_2}
\end{eqnarray}
where (compare to Eq. \ref{eq:ManyBodyPerturbationTheoryInExactFactorizationEq_7})
\begin{equation} 
\vec{C}'^{\left(2-l_{\bar{s}}\right)}_{mk_{\bar{s}}}\left(\vec{R}',\bar{z}_{\bar{s}}\right) \equiv \chi^{\ast}_{m}\left(\vec{R}'\right) \vec{P}'^{2-l_{\bar{s}}}_{k_{\bar{s}}}\left(\bar{z}_{\bar{s}}\right) \chi_{m}\left(\vec{R}'\right).
\label{eq:PerturbationExpansionInTheBodyFixedFrameEq_3}
\end{equation}
For the sake of notational simplicity, we used the same notation for the exact factorized body-fixed states as we did for the laboratory frame exact factorized states. The terms in the Hamiltonian we start with, $H_{b} \approx T'_{n} + H'_{BO}$, are defined in Eqs. 16-18 of Ref. \cite{Harkonen-ManybodyGreensFunctionTheoryOfElectronsAndNucleiBeyondTheBornOppenheimerApproximation-PhysRevB.101.235153-2020}. The expansions in Eqs. \ref{eq:PerturbationExpansionInTheBodyFixedFrameEq_1} and \ref{eq:PerturbationExpansionInTheBodyFixedFrameEq_2} allows us to write the exact electronic Green's function $G'\left(\vec{y}t,\vec{y}'t'\right)$ given by Eq. \ref{eq:BeyondBornOppenheimerEq_10} in terms of exact factorized states. The simplest approximation is the BO approximation (in the body fixed frame) and these results provide a systematic way to go beyond it.

Next we consider the nuclear Green's functions in the body-fixed frame. The expansion of the nuclear Green's function given by Eq. \ref{eq:BeyondBornOppenheimerEq_11} can be written in terms of the relations (see Eqs. \ref{eq:ManyBodyPerturbationTheoryInExactFactorizationEq_8} and \ref{eq:ManyBodyPerturbationTheoryInExactFactorizationEq_9})
\begin{eqnarray} 
&&\text{Tr}\left[ \mathcal{T}_{c} \left\{ e^{ -i \int_{c}d\bar{z} \hat{H}''_{n}\left(\bar{z}\right) } \hat{u}'_{\alpha_{\bar{n}}}\left(k_{\bar{n}}z_{\bar{n}}\right) \right\} \right]_{\Psi_{b}} \nonumber \\
&&\ \ \ = \sum_{m} \int d\vec{r}' \int d\vec{R}' \sum^{\infty}_{s = 0} \frac{\left(-i\right)^{s}}{s!} \int_{c} d\bar{z}_{\bar{s}} \nonumber \\
&&\ \ \ \times \sum_{k_{\bar{s}}} \frac{ 1 }{2^{s} M_{k_{\bar{s}}}} \sum^{2}_{l_{\bar{s}} = 0} \binom{2}{l_{\bar{s}}} \Phi^{\left(m\right)\ast}_{\vec{R}'}\left(\vec{r}'\right) \chi^{\ast}_{m}\left(\vec{R}'\right) \nonumber \\
&&\ \ \ \times \mathcal{T}_{c} \left\{ e^{ -i \int_{c}d\bar{z} \left[\epsilon'_{m}\left(\bar{z}\right) - U'^{\left(m\right)}_{en}\left(\bar{z}\right)\right] } \hat{u}'_{\alpha_{\bar{n}}}\left(k_{\bar{n}}z_{\bar{n}}\right) \right. \nonumber \\
&&\ \ \ \times \left. \left[\vec{P}'^{l_{\bar{s}}}_{k_{\bar{s}}}\left(\bar{z}_{\bar{s}}\right) \Phi^{\left(m\right)}_{\vec{R}'}\left(\vec{r}'\right)\right] \left[ \vec{P}'^{2-l_{\bar{s}}}_{k_{\bar{s}}}\left(\bar{z}_{\bar{s}}\right) \chi_{m}\left(\vec{R}'\right)\right] \right\}, \nonumber \\
\label{eq:PerturbationExpansionInTheBodyFixedFrameEq_4}
\end{eqnarray}
and
\begin{eqnarray} 
&&\text{Tr}\left[ \mathcal{T}_{c}\left\{ e^{ -i \int_{c}d\bar{z} \hat{H}''_{n}\left(\bar{z}\right) } \right\} \right]_{\Psi_{b}} \nonumber \\
&&\ \ \ = \sum_{m} \int d\vec{r}' \int d\vec{R}' \sum^{\infty}_{s = 0} \frac{\left(-i\right)^{s}}{s!} \int_{c} d\bar{z}_{\bar{s}} \nonumber \\
&&\ \ \ \times \sum_{k_{\bar{s}}} \frac{ 1 }{2^{s} M_{k_{\bar{s}}}} \sum^{2}_{l_{\bar{s}} = 0} \binom{2}{l_{\bar{s}}} \Phi^{\left(m\right)\ast}_{\vec{R}'}\left(\vec{r}'\right) \chi^{\ast}_{m}\left(\vec{R}'\right) \nonumber \\
&&\ \ \ \times \mathcal{T}_{c}\left\{ e^{ -i \int_{c} d\bar{z} \left[\epsilon'_{m}\left(\bar{z}\right) - U'^{\left(m\right)}_{en}\left(\bar{z}\right)\right] } \left[\vec{P}'^{l_{\bar{s}}}_{k_{\bar{s}}}\left(\bar{z}_{\bar{s}}\right) \Phi^{\left(m\right)}_{\vec{R}'}\left(\vec{r}'\right)\right] \right. \nonumber \\
&&\ \ \ \times \left. \left[ \vec{P}'^{2-l_{\bar{s}}}_{k_{\bar{s}}}\left(\bar{z}_{\bar{s}}\right) \chi_{m}\left(\vec{R}'\right)\right] \right\}.
\label{eq:PerturbationExpansionInTheBodyFixedFrameEq_5}
\end{eqnarray}
In Eqs. \ref{eq:PerturbationExpansionInTheBodyFixedFrameEq_4} and \ref{eq:PerturbationExpansionInTheBodyFixedFrameEq_5}, the following Hamiltonian appears
\begin{equation} 
\hat{H}''_{n}\left(\bar{z}\right) \equiv \hat{T}'_{n}\left(\bar{z}\right) + \hat{\epsilon}'\left(\bar{z}\right) - \hat{U}'_{en}\left(\bar{z}\right),
\label{eq:PerturbationExpansionInTheBodyFixedFrameEq_6}
\end{equation}
where all the terms are written in terms of body-fixed variables. The results given by Eqs. \ref{eq:PerturbationExpansionInTheBodyFixedFrameEq_4} and \ref{eq:PerturbationExpansionInTheBodyFixedFrameEq_5} allow us to write the nuclear Green's function defined by Eq. \ref{eq:BeyondBornOppenheimerEq_11} in terms of the exact factorized state. This in turn makes possible to derive beyond-BO approximations systematically. The Hamiltonian of Eq. \ref{eq:PerturbationExpansionInTheBodyFixedFrameEq_6} is a generalization of the phonon Hamiltonian in the BO approximation as the phonon Hamiltonian is included to $\hat{H}''_{n}$. There are some additional potential terms appearing in $\hat{H}''_{n}$ included to the operators $\hat{\epsilon}'$ and $\hat{U}'_{en}$ which originate from the nuclear kinetic energy. These terms introduce correction terms to the BO potential and have an effect on the nuclear properties, like the phonon spectrum of the system, and therefore to the electronic properties as the electron and nuclear systems are coupled.

The Green's functions obtainable from Eqs. \ref{eq:PerturbationExpansionInTheBodyFixedFrameEq_1}, \ref{eq:PerturbationExpansionInTheBodyFixedFrameEq_2}, \ref{eq:PerturbationExpansionInTheBodyFixedFrameEq_4} and \ref{eq:PerturbationExpansionInTheBodyFixedFrameEq_5} form a Green's function theory generalizing the BO approach. The dynamics for these functions will be rather complicated, but we believe that the lowest order approximations to these functions provide a convenient tool to study beyond-BO physics in various systems of relevance.

\begin{acknowledgments}
The author gratefully acknowledge funding from the Magnus Ehrnrooth and Wihuri Foundations.
\end{acknowledgments}

\bibliography{bibfile}
\end{document}